\documentclass[prl,twocolumn,amsmath,amssymb,floatfix,showpacs,longbibliography]{revtex4-1}

\usepackage{tikz}
\usepackage{graphicx}
\usepackage{dcolumn}
\usepackage{bm}
\usepackage{mciteplus}
\usepackage{braket}

\usepackage[english]{babel}

\usepackage[size=small]{caption}
\usepackage{etoolbox}

\usepackage[normalem]{ulem}

\newcommand{\beq}{\begin{equation}}
\newcommand{\eeq}{\end{equation}}
\newcommand{\MB}{\langle \bf B \rangle} 
\newcommand{\bs}{\tilde{\bf b}}

\DeclareCaptionFormat{algorithms}{\vskip-15pt\hrulefill\par#1#2#3\vskip-6pt\hrulefill}
\usepackage{shortCut}

\begin{document}

\title{The fate of alpha dynamos at large $Rm$}
\author{Alexandre \textsc{Cameron}}
\email[]{alexandre.cameron@ens.fr}
\author{Alexandros \textsc{Alexakis}} 
\email[]{alexakis@lps.ens.fr}
\affiliation{Laboratoire de Physique Statistique, 
  \'Ecole Normale Sup\'erieure, PSL Research University; 
  Universit\'e Paris Diderot Sorbonne Paris-Cit\'e; 
  Sorbonne Universit\'es UPMC Univ Paris 06; CNRS; 
  24 rue Lhomond, 75005 Paris, France}
\date{\today}
\begin{abstract}
At the heart of today's solar magnetic field evolution models lies the alpha dynamo description. 
In this work, we investigate the fate of alpha-dynamos as the magnetic Reynolds number $Rm$ is increased.
Using Floquet theory, we are able to precisely quantify mean field effects like the alpha and beta effect 
(i) by rigorously distinguishing dynamo modes that involve large scale components from the ones that only involve small scales, and by 
(ii) providing a way to investigate arbitrary large scale separations with minimal computational cost.
We apply this framework to helical and non-helical flows as well as to random flows with short correlation time.
Our results determine that the alpha-description is valid for $Rm$ smaller than a critical value $Rm_c$ at which small scale dynamo instability starts.
When $Rm$ is above $Rm_c$ the dynamo ceases to follow the mean field description and
the growth rate of the large scale modes becomes independent of the scale separation 
while the energy in the large scale modes is inversely proportional to the square of the scale separation. 
The results in this second regime do not depend on the presence of helicity.
Thus alpha-type modeling for solar and stellar models needs to be reevaluated 
and new directions for mean field modeling are proposed.

\end{abstract}
\maketitle

\section{Introduction}        
\label{sec:intro}             

Dynamo instability refers to the spontaneous amplification of magnetic energy
due to the stretching and refolding of magnetic field lines by a flow.
It explains the presence of magnetic fields through out 
the universe from planetary to galactic scales. 
In many of these cases, dynamo action produces ordered fields 
of scale $L$ much larger than the typical underlying turbulent scales $\ell$. 
A prominent example is the sun whose magnetic field possesses a 
time and spatial coherence much larger than the typical turbulent time and length 
scales \cite{Charbonneau2014Solar,Rieutord2008Solar,weiss2008solar}. 
A mechanism for the generation of such large scale magnetic fields by small 
scale turbulent eddies was proposed by E. Parker in \cite{parker1955hydromagnetic},
where he considered the evolution of large scale fields due to the averaged effect 
of small scale eddies that lack parity invariance.
This idea has led to the concept of mean-field magneto-hydrodynamics \cite{steenbeck1966berechnung,childress1969class,Moffatt1978book,Krause1980Mean} 
where the averaged effect of small scale velocity field is taken into account through the calculation of transport coefficients.

The starting point for these calculations is the magnetic induction equation for the magnetic field $\bf B$:
\beq \partial_t {\bf B = \nabla \times ( u \times B) + \eta \nabla^2 B} \label{induction} \eeq  
that is advected by a small scale velocity $\bf u$ under the effect of magnetic diffusion $\eta$.
The magnetic field is then split in a mean part $\MB$ (averaged over the small scales) and a fluctuating part $\bf b$
so that $\bf B=\MB+b$ and $\langle \bf b \rangle=0$.
The averaged equation for the large scale magnetic field reads:
\beq \partial_t {\bf \MB = \nabla \times \mathcal{E} + \eta \nabla^2 \MB } \label{MB} \eeq    
where the mean electromotive force $\mathcal{E}=\bf \langle u \times b \rangle $
is a measure of the cross correlation of the small scale velocity $\bf u$ and magnetic $\bf b$ fields.
It can be found by solving for the evolution of the small scale field $\bf b$: 
\beq \partial_t {\bf b - \eta \nabla^2 b = \nabla \times ( u \times \MB) + \nabla \times G } \eeq     
where $\bf G = u \times b - \langle u \times b \rangle$.
If $\bf G$ can be neglected (which implies that there is no small scale dynamo) $\bf b$ has a linear dependence on $\MB$
that acts as a source term for the small scale fluctuations.
In this case the mean electromotive force can be expanded in a series of the gradients of the large scale magnetic field as:
\beq \mathcal{E}^i = \alpha^{ij} \langle B \rangle^j + \beta^{ijk} \nabla^j \langle B\rangle^k + \dots \,. \label{TC} \eeq          
The tensors $\alpha,\beta,\dots$ are transport coefficients that depend on the properties of the small scale velocity field. 
In particular the first tensor $\alpha$ is non-zero if the flow is helical.
It can drive large scale magnetic field amplification with a growth rate $\gamma$ that is proportional to the scale separation $\gamma\propto \ell/L$.
These type of dynamos are referred to as alpha-dynamos in the literature.
In the absence of helicity, large scale dynamos are also known to exist through an instability related to the second tensor $\beta$ \cite{Lanotte1999}.
This effect leads to a growth rate proportional to the square of the scale separation $(\ell/L)^2$.
Both cases are examples of large scale dynamos (LSD).

Given the value of these tensors and inserting eq.~\eqref{TC} in eq.~\eqref{MB} one obtains a closed equation for the large scale magnetic field.
This allows to compute the large scale evolution without knowing the precise details of small scale turbulence.
This procedure is commonly used in solar \cite{brun2015recent,dikpati2006simulating,choudhuri2007predicting} 
and planetary models \cite{Schrinner2005Mean}. Due to limited computational power, these models only 
compute the large scale magnetic field 
while the effect of small scale fluctuations is modeled through the transport coefficients. 
If these coefficients are properly parametrized these models reproduce the observed behavior of the solar magnetic field. 
Global models that solve the full stellar system without parametrization  still fall short 
of reproducing quantitatively the solar cycle despite the great advancement in recent calculations 
\cite{glatzmaier1985numerical,gilman1983dynamically,brun2004global,browning2006dynamo,ghizaru2010magnetic,miesch2009turbulence}.

However calculating the transport coefficients from first principles remains non-trivial.
It can be achieved when the magnetic Reynolds number: $Rm=U\ell/\eta$ (where $U$ is the rms value of the velocity field)
is much smaller than unity $Rm\ll1$. In this case, the small scale induction equation can 
be simplified to $\eta \nabla^2 \bf b = -\nabla \times ( u \times \MB)$ and easily solved by spectral methods \cite{steenbeck1966berechnung}. 
Another frequent approximation consists in assuming that the velocity field has a very short correlation time $\tau$ compared to 
the eddy turnover time (see discussion in \cite{Radler2007Meanfield}). The solution is then approximated to $\bf b \approx \tau \nabla \times ( u \times \MB)$. 
Both cases lead to a linear dependence of $\bf b$ on $\MB$ in agreement with $\alpha$-modeling,
and lead to a non-zero alpha-effect provided that the flow is helical.
In particular for the small $Rm$, the $\alpha$-tensor can be rigorously calculated using multi-scale methods \cite{childress1969class}.
However, for natural flows, neither of these assumptions hold and different methods have been devised 
to measure the transport coefficients by numerical simulations of small scale turbulence
\cite{schrinner2007mean,Rheinhardt2010Test,Brandenburg2008Kinematic,brandenburg2008scale}.

For large values of $Rm$, which correspond to astrophysical regimes, neglecting the $\bf G$ term is not 
necessarily a valid assumption. Indeed, at sufficiently large $Rm$, small scale dynamo (SSD) action is expected 
to take place and small scale magnetic fields to be self-generated, exponentially amplifying the value of the electromotive force.
This is against the basic assumption made above that the electromotive force has a linear dependence on the large scale field $\MB$.
Indeed, many authors have questioned the validity of alpha modeling beyond the critical value of $Rm_c$ where 
SSD takes place \cite{Boldyrev2005Magnetic,Courvoisier2006alpha,Hughes2008mean,Cattaneo2009Problems,cattaneo2014large}.
 
Part of their objections can be elegantly summed up in the following two-mode model.
We consider the evolution of a large scale mode $b_q$ at wave number $q\propto1/L$ and a small scale mode $b_k$ 
at wavenumber $k\propto1/\ell$, with $q\ll k$, that are coupled by an alpha effect as follows:
\begin{equation}
\begin{array}{lrlrl}
	\dot{b}_q =& -\eta    q^2 &  b_q    & +  \alpha   q                 & b_k \, ,\\
  \dot{b}_k =&  \alpha  k   &  b_q    & +  \gamma_{_{SSD}}                & b_k \, 
\end{array}
\end{equation}
where $\gamma_{_{SSD}}=u_k k -\eta k^2$ is the growth rate of the SSD obtained by setting $\alpha=0$.
It is positive if $Rm=u_k/\eta k >1$ that marks the SSD onset. 
Looking for exponential solutions $(b_q,b_k)\propto e^{\gamma t}$ the growth rate $\gamma$ of the two modes can be explicitly calculated and it is given by
$\gamma = \frac{1}{2}\left[ \gamma_{_{SSD}} -\nu q^2 \pm \sqrt{ \gamma_{_{SSD}}^2 + 4 \alpha kq  + 2\gamma_{_{SSD}} \eta q^2 +\eta^2 q^4  } \right]$.  
One notices directly that in the $q\ll k$ limit, if $\gamma_{_{SSD}}<0$, the system has one negative eigenvalue $\gamma\simeq \gamma_{_{SSD}}$ 
and one positive eigenvalue $\gamma \simeq \alpha^2 kq/|\gamma_{_{SSD}}| $. The growing eigen mode satisfies 
$b_q/b_k \simeq (|\gamma_{_{SSD}}|/\alpha k) =\mathcal{O}(1)$.
On the other hand, if $\gamma_{_{SSD}}>0$, the system has one positive eigenvalue $\gamma\simeq \gamma_{_{SSD}}$ and its eigenvector satisfies 
$b_q/b_k  \simeq (\alpha q/|\gamma_{_{SSD}}|)=\mathcal{O}(q/k) $. Thus, beyond the SSD dynamo onset, the growth rate does not satisfy 
the scaling $\gamma\propto q$ while the projection of the unstable eigenmode on the large scales decreases with scale separation.

To demonstrate the above arguments and the possible failure of the LSD description, the notion of 
scale separation needs to be clearly formulated. This has been attempted in the past using direct
numerical simulations \cite{ponty2011transition} but only for moderate scale separations. 
A precise way to quantify the evolution of large scales can be done using Floquet theory \cite{floquet_sur_1883}
also known as Bloch theory in quantum mechanics \cite{Bloch1929}. Floquet theory can be applied to the
linear evolution of the magnetic field ${\bf B}({\bf x} ,t)$ driven by a spatially periodic 
flow ${\bf u}({\bf x},t)$ of a given spatial period $\ell=2\pi/k$. 
Under these assumptions Floquet theory states that the magnetic field can be decomposed as ${\bf B}({\bf x} ,t)=e^{i\bf q\cdot x} \bs({\bf x} ,t)+c.c.$ 
where ${\bs}({\bf x} ,t)$ is a complex vector field that satisfies the same spatial periodicity as the velocity field $\bf u$, and $\bf q$ is an arbitrary wave number. 
For $q=|{\bf q}|\ll k $, the volume average $\langle \bs \rangle$ over one spatial period $(2\pi/k)^3$
gives the amplitude of $\bs$ at large scales $L\propto 1/{ q}$. 
Thus, fields with $q=0$ and $\langle \bs \rangle=0$ correspond to purely small scale fields.
If such fields are dynamo-unstable, the system has a SSD instability and  we will denote its growth rate as $\gamma_{_{SSD}}$.
For $0< q < 1$  the dynamo mode has in general a finite projection to the large scales measured by $\langle \bs \rangle$.
%
Substituting in the induction eq.~\eqref{induction}, we obtain:
\beq \partial_t {\bs}  =i {\bf q } \times ( {\bf u } \times \bs) + \nabla \times ( {\bf u} \times \bs)  + \eta (\nabla+i {\bf q})^2 \bs \label{floquet} \eeq    
Note that now $\bf q$ is a control parameter that can be taken to be arbitrarily small. 
The gain in using the Floquet framework is twofold:
(i) it provides us with a clear way to disentangle dynamos that involve only small scales (for which ${\bf q}/k \in \mathbb{Z}^3$) 
    from dynamos that involve large scales ($0<q/k\ll 1$);
(ii) it allows to investigate numerically arbitrary large scale separations ${q}\ll k $ with no additional numerical cost.

\begin{figure}[!ht]
   \centering
   \includegraphics[width=\columnwidth ]{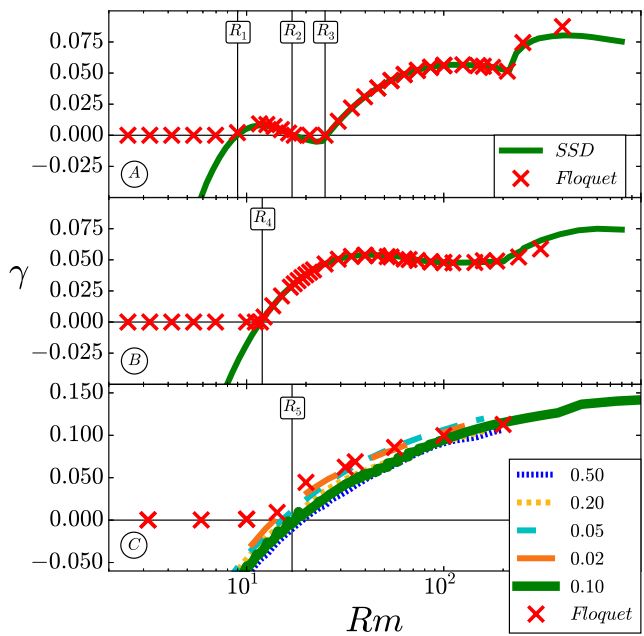}
   \caption{Growth rate as a function of $Rm$ for the different flows considered.
   The SSD results are given by the solid lines, while the results from the Floquet code with $q=10^{-3}$ are denoted by crosses. 
   In the bottom panel, the value of $\tau=0.1$ was used for the Floquet code, and different values of $\tau$ were used 
   for the SSD as indicated. }
   \label{fig1}
 \end{figure}

In this work, we consider the velocity fields parametrized as:
\beq
{\bf u}= U\, \left[
\begin{array}{ccc}
\sin(k y + \phi_2 ) &+& \cos(k z + \psi_3) , \\ \sin(k z + \phi_3 ) &+& \cos(k x + \psi_1), \\ \sin(k x + \phi_1 ) &+& \cos(k y + \psi_2)
\end{array}
 \right].
\eeq
Three cases are examined.
In the first case ($A$), $\phi_i=\psi_i=0$ for all $i\in(1,2,3)$, the flow corresponds to the well studied helical 
ABC flow \cite{Gallaway1984,Galloway1986,galanti1992linear,Alexakis2011,Bouya2013,Jones2014}, in the second case ($B$), $\phi_i=\psi_i-\pi/2=0$ is a non-helical flow, and in the
last case ($C$) the phases $\phi_i=\psi_i$ change randomly every time $\tau$ and corresponds to random helical flow.
For the time-independent flows (cases $A$ and $B$), the magnetic Reynolds number is defined as $Rm=U/k\eta$ and the growth rate is measured in units of $Uk$.
For the random flow (case $C$), the definition $Rm=(U/k\eta)\times (\tau Uk)= U^2\tau /\eta$ is used and the growth rate is measured in units of $U^2k^2\tau$.
The latter definition takes into account
that a fast de-correlation time reduces the rate at which the flow shears the magnetic field lines. 
As will be shown in fig.~\ref{fig1}, this scaling makes the results collapse on the same curve small $\tau$.
Eq.~\ref{floquet} was solved numerically and the dynamo growth rate $\gamma$ was measured for various values of $Rm$ and ${\bf q}=\hat z q$
using a pseudospectral code in a cubic periodic domain of side $2\pi$ with $k=1$ and spatial resolution ranging from $32^3$ to $128^3$ depending on $Rm$.
Details on the Floquet code can be found in \cite{cameron2016large}.
The results are compared with the SSD growth rate $\gamma_{_{SSD}}$ obtained from a tested dynamo code \cite{mininni_hybrid_2011}.

The calculated growth rates are plotted in fig.~\ref{fig1} as a function of the Reynolds number 
for the three different velocity fields used. 
Crosses correspond to the results obtained from the Floquet code with $q=10^{-3}$ while $\gamma_{_{SSD}}$ is shown with a solid green line. 
In the first flow (A) the $\gamma_{_{SSD}}$ reproduces the classical `two-window' result 
of the ABC dynamo \cite{Gallaway1984,Galloway1986,galanti1992linear,Alexakis2011,Bouya2013,Jones2014} for which SSD
exists for $Rm$ in the range $R_1<Rm<R_2$ and $R_3<Rm$. 
For the non-helical case (B) SSD appears for $Rm>R_4\simeq 12$.
In the case (C), different values of $\tau$ were used in the range $(0.02,0.5)$ as mentioned in the legend. 
For the Floquet results the value of $\tau$ used was $\tau=0.1$. 
SSD appears above a critical value $R_5$ that weakly depends on the value of $\tau$. 
At sufficiently small $\tau$, the critical value of $Rm=R_5$, at which SSD appears, becomes independent of $\tau$ with $R_5\simeq 15$.
All three cases show the same feature: 
when $\gamma_{_{SSD}}>0$, the Floquet and SSD results have the same growth rate, while, when $\gamma_{_{SSD}}<0$, 
the Floquet results have a positive growth rate but of order $q$ (or $q^2$).  
 
\begin{figure}[!ht]
    \centering
    \begin{tikzpicture}
        \node[anchor=south west,inner sep=0] (image) at (0,0) 
        {\includegraphics[width=\columnwidth,trim= 0 0 0 0, clip=true]{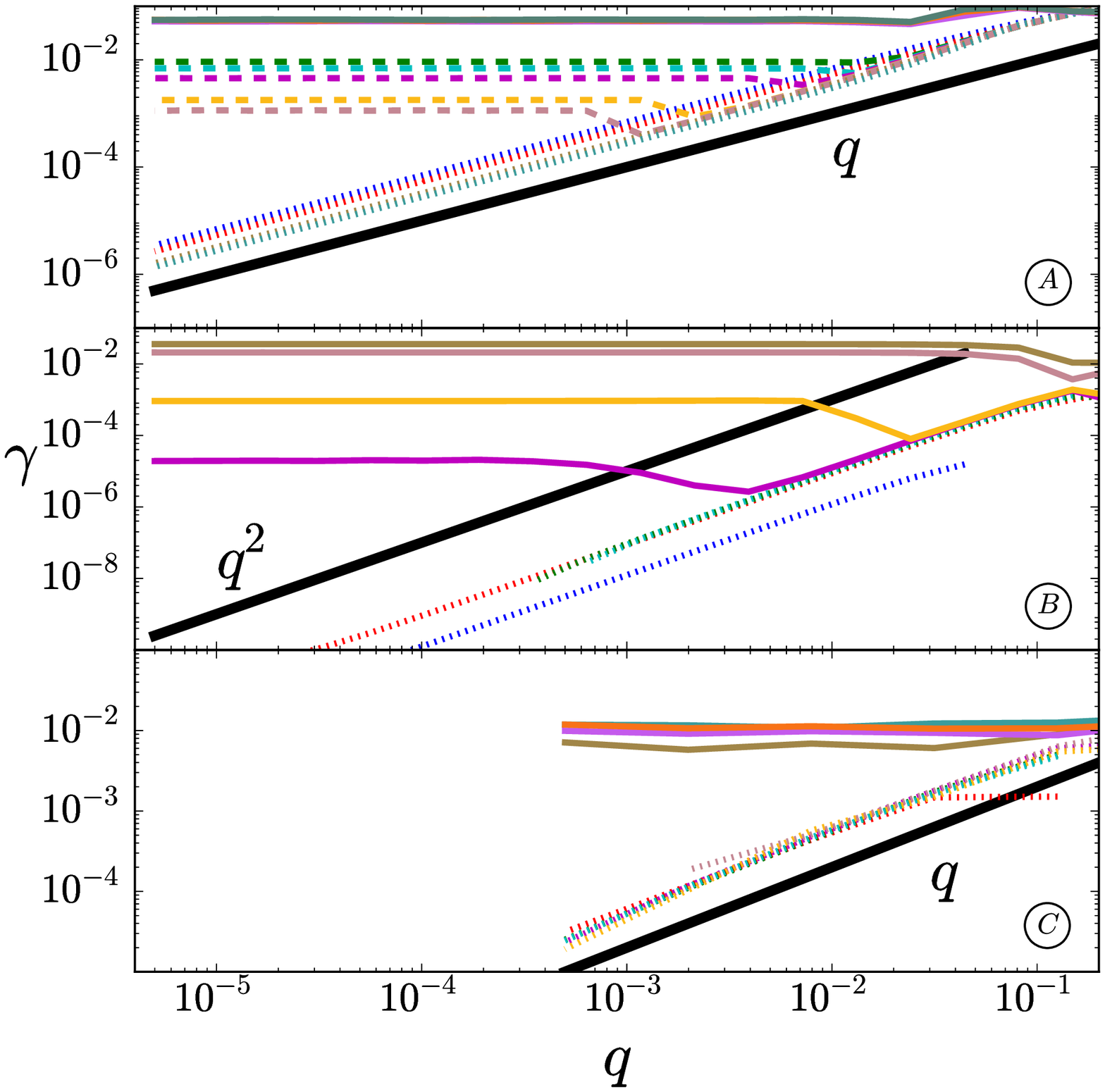}};
        \begin{scope}[x={(image.south east)},y={(image.north west)}]
            \node[anchor=south west,inner sep=0] (image) at (.135,0.125) 
            {
			 \includegraphics[width=3.cm,trim= 5 10 5 4, clip=true]{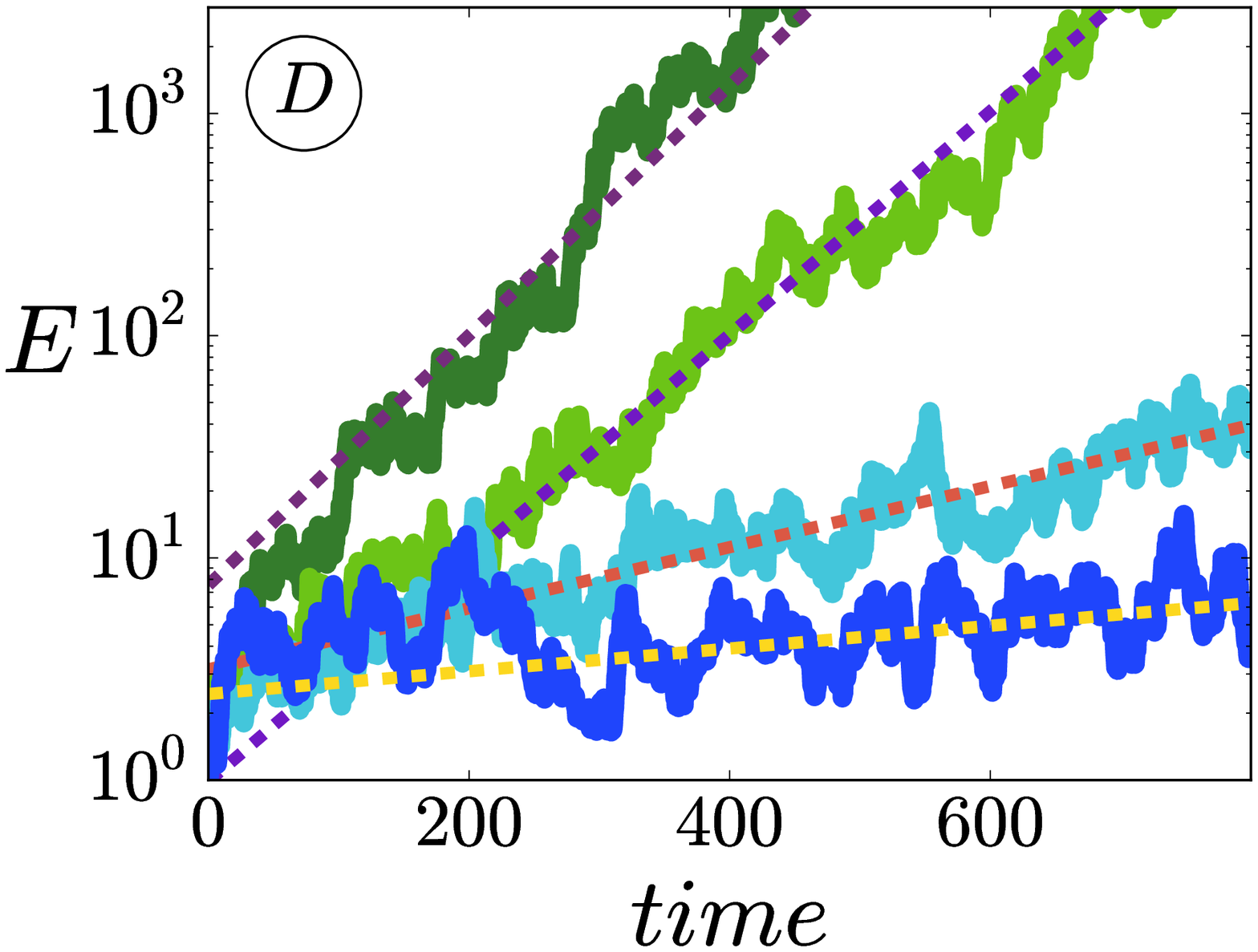}            
};
        \end{scope}
    \end{tikzpicture}
  \caption{The growth rate as a function of $q$ for different values of $Rm$. The line types are as follows. 
           Panel (A): For $Rm<R_1$ and  $R_2<Rm<R_3$ (dotted lines), 
                      for $R_1<Rm<R_2$ (dashed lines),
                      for $R_3<Rm$ (solid lines).
           Panel (B): For $Rm<R_4$ (dotted lines), 
                      for $R_4<Rm$ (solid lines).
           Panel (C): For $Rm<R_5$ (dotted lines), 
                      for $R_5<Rm$ (solid lines).
                      The inset (D) shows a typical signal for the evolution of energy from case (C) for $Rm<R_5$.}
  \label{fig2}
\end{figure}

This observation is achieved examining the dependence of the growth rate on $q$ shown in the three panels of fig.~\ref{fig2}. 
For each line in these figures, a series of simulations of fixed $Rm$ and varying $q$ was performed.
Each line corresponds to a different value of $Rm$.
Panel (A) shows the growth rate for the ABC flow. 
For the values of $Rm<R_1$ and $R_2<Rm<Rm_3$ (where there is no SSD), the growth rate is plotted with dotted lines;
the first dynamo window $R_1<Rm<Rm_2$ is plotted using dashed lines;
while in the range $R_3<Rm$ solid lines are used.
It is clear that for the no-small-scale-dynamo range
a $\gamma \propto q$ scaling is followed (alpha dynamos) 
while in the presence of SSD $\gamma$ is independent of the value of $q$.
Similarly, for the non-helical runs, in the absence of SSD ($Rm<R_4$ dotted lines), the growth-rate follows the scaling: $\gamma\propto q^2$, which indicates a $\beta$-type dynamo instability,
while in the presence of SSD (solid lines) there is no dependence of the growth-rate on $q$.
Even in the random flow, the same feature is observed: for $Rm<R_5$ (dotted lines) the results show a $\gamma \propto q$ scaling 
that suggest the presence of a random $\alpha$-effect,
but for $Rm>R_5$ this behavior transitions to a $q$-independent growth-rate (solid lines).
We note that due to the random nature of this flow the accuracy of our measurements is limited and we only
examine values of $q>5\cdot10^{-4}$. The inset (D) shows a typical signal for the evolution of energy from case (C).

\begin{figure}[!ht]
  \centering
      \begin{tikzpicture}
        \node[anchor=south west,inner sep=0] (image) at (0,0) 
        {\includegraphics[width=\columnwidth,trim= 0 0 0 0, clip=true]{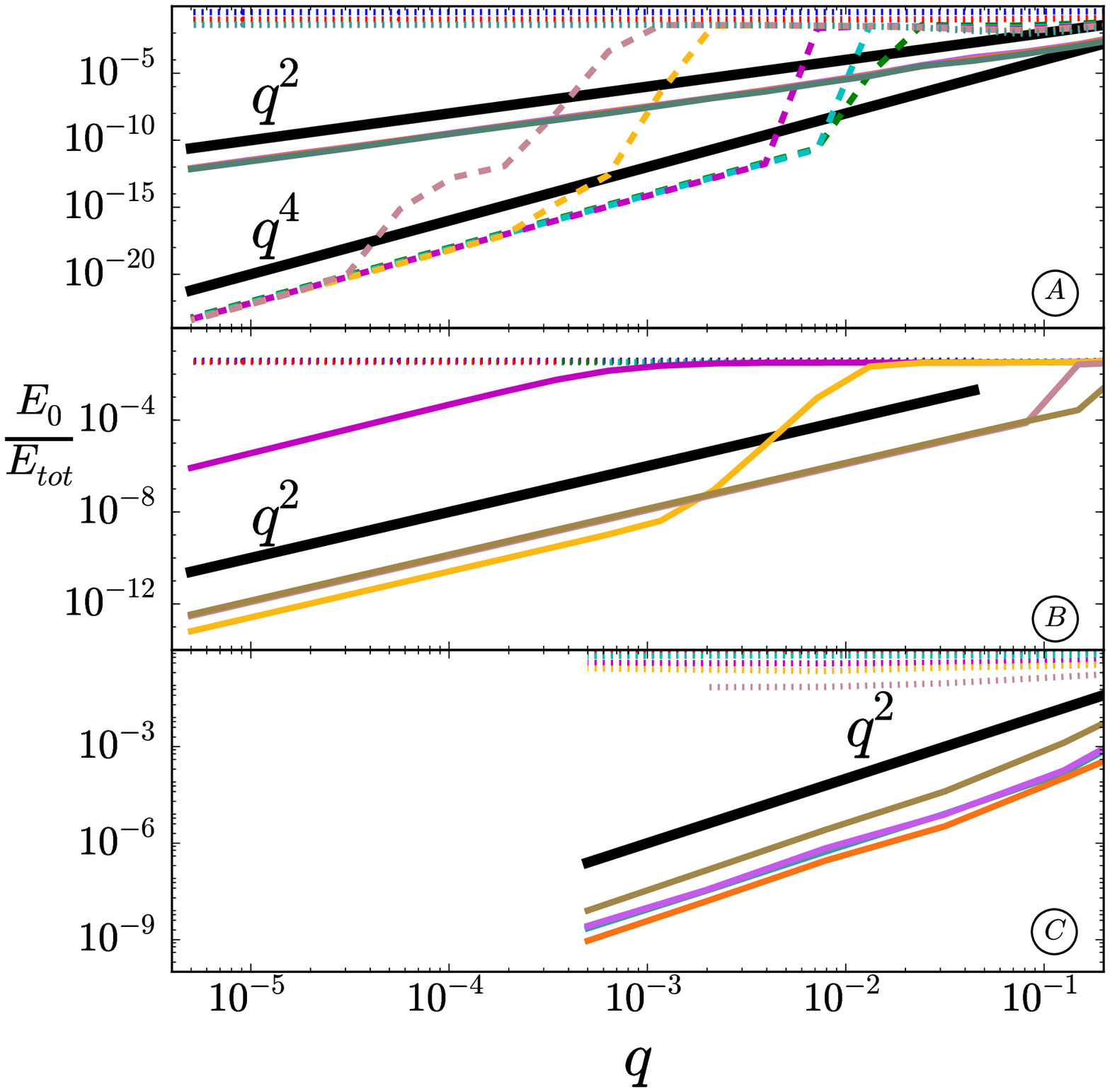}};
        \begin{scope}[x={(image.south east)},y={(image.north west)}]
            \node[anchor=south west,inner sep=0] (image) at (.17,0.125) 
            {
			 \includegraphics[width=3.cm,trim= 5 10 5 4, clip=true]{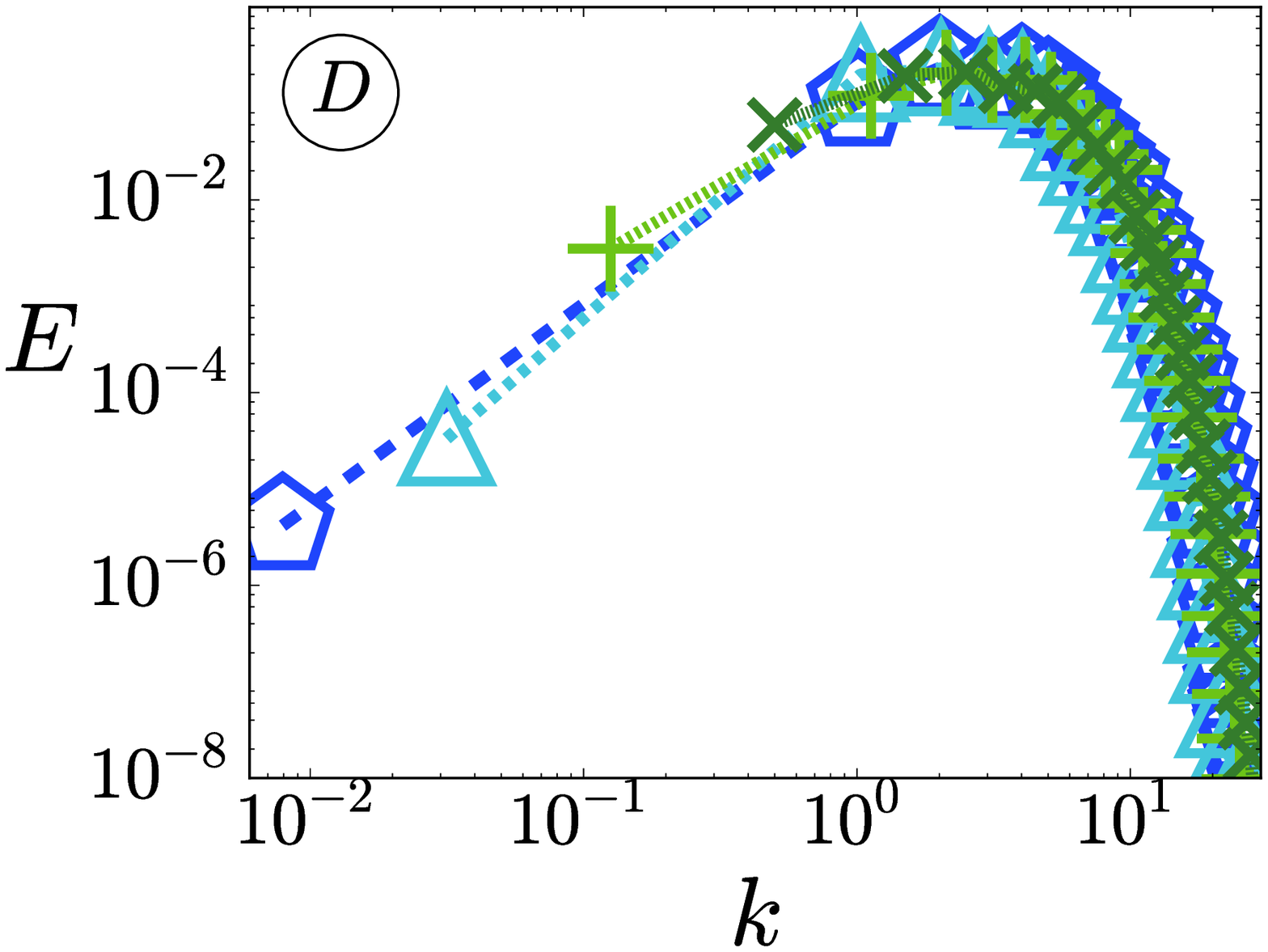}    
};
        \end{scope}
    \end{tikzpicture}
  \caption{The energy ratio $E_0/E_{tot}$. The line styles are as in figure \ref{fig2}. 
  The inset (D) shows the energy spectra for four different values of $q$ of case (C) at the highest $Rm$.}
  \label{fig3}
\end{figure}

At first, a finite growth-rate $\gamma>0$ in the limit of $q\to0$ seems to violate the flux conservation. 
Indeed, flux conservation enforces modes with $q = 0$, corresponding to uniform fields, not to grow.
The explanation is found by looking at the projection of the unstable modes to the large scales. 
In fig.~\ref{fig3}, we plot the ratio of the energy contained in the large scale mode $e^{i\bf q\cdot x}$ that is given by $E_0=\frac{1}{2} |\langle \bs \rangle |^2$
to the total energy $E_{tot}=\frac{1}{2} \langle {|\bs|}^2 \rangle$ as a function of $q$ for the same values of $Rm$ as used in fig.~\ref{fig2} and the same line types.
For LSD (of the type $\alpha$ or $\beta$) the projection to the large scales
becomes independent of $q$ for $q\to 0$ (although it still depends on the value of $Rm$). As $Rm$ approaches the SSD onset, 
this projection decreases. For values of $Rm$ larger than the onset of the SSD, the projection to the large scale modes 
becomes dependent on $q$ and follows the scaling $\gamma \propto q^2$ in most cases or $\gamma \propto q^4$ for the case of the first dynamo window in the ABC flow. 
This result can be obtained by a regular expansion of eq.~\eqref{floquet} for small $q$ such that $\gamma=\gamma_0 + q \gamma_1 +\dots$ and
${\bs}={\bs}_0 + q {\bs}_1 \dots$. At zeroth order, one obtains $\gamma=\gamma_{_{SSD}}$ and $\langle {\bs}_0 \rangle=0$. 
At next order, by averaging over space, one obtains $\gamma_0 \langle {\bs}_1 \rangle = i {\bf q} \times \langle {\bf u} \times \bs_0 \rangle$.
This last result shows that the energy in the large scale mode scales like $q^2$, provided that the mean electromotive force $\langle {\bf u} \times \bs_0 \rangle$
due to the SSD mode is not zero. If it is zero, then the next order term leads to a $q^4$ scaling and so on.
Note that this argument does not depend on the presence or absence of helicity in the flow.
In fact as shown in the top panel of fig.~\ref{fig3}, the same flow results in different scalings of $E_0/E_{tot}$ depending on which dynamo window is examined.
Indeed, in the first window $R_1<Rm<R_2$, the most unstable mode possesses different symmetries than the most unstable mode for $Rm>R_3$ \cite{Jones2014}.

The results above give a clear description of the transition from SSD to LSD. 
Below the SSD onset, the mean field predictions are valid and lead to a growth rate proportional to $q$ or $q^2$ depending 
whether an $\alpha$- of $\beta$-dynamo is present. Above the SSD onset large scales grow with the $\gamma_{_{SSD}}$ growth rate 
but with a projection to the large scales that decreases with a scale separation. This behavior cannot be modeled
with terms that are linear in the amplitude of the large scale field as eq.~\eqref{TC} implies.
On the contrary, the behavior of the large scales mode depends on SSD.
Despite its small projection, it has a faster growth rate than mean field dynamos.
Therefore, the large scales mode could possibly be modeled as a non-homogeneous term
in the mean field dynamo equation.
This possibility however requires further investigations.

\acknowledgments 

This work was granted access to the HPC resources of MesoPSL financed by the Region 
Ile de France and the project Equip@Meso (reference ANR-10-EQPX-29-01) of 
the programme Investissements d'Avenir supervised by the Agence Nationale pour 
la Recherche and the HPC resources of GENCI-TGCC-CURIE \& GENCI-CINES-JADE 
(Project No. x20162a7620) where the present numerical simulations have been performed.

\bibliography{alpha}

\end{document}